\documentclass{aa}
\usepackage{graphicx,epsfig}
\def\msun{{\rm M_\odot}}
\def\msuny{{\rm M_\odot y^{-1}}}
\def\cm{{\rm cm}}
\def\K{{\rm K}}
\def\g{{\rm g}}
\def\s{{\rm s}}
\def\km{{\rm km}}
\newcommand{\chem}[2]{$\rm{}^{#1}\kern-0.8pt#2$}
\newcommand{\chim}[2]{\rm{}^{#1}\kern-0.8pt#2}
\newcommand{\reac}[6]{$\rm\,{}^{#1}\kern-0.8pt{#2}\,({#3}\,,{#4})\,
           {}^{#5}\kern-0.8pt{#6}\,$}

\begin{document}

   \title{He-detonation in sub-Chandrasekhar CO white dwarfs: a new insight into
	  energetics and p-process nucleosynthesis}

   \author{Stephane Goriely
           \inst{1},
           Jordi Jos\'e
           \inst{2},
           Margarita Hernanz
           \inst{3},
           Marc Rayet\inst{1}
           \and
           Marcel Arnould\inst{1}
          }

   \offprints{S. Goriely}

   \institute{Institute d'Astronomie et d'Astrophysique, Universit\'e
              Libre de Bruxelles, CP 226, B-1050 Brussels, Belgium.
         \and
              Departament de Fisica i Enginyeria Nuclear, Universitat
              Politecnica de Catalunya, Av. V\'\i ctor Balaguer s/n,
              E-08800 Vilanova i la Geltr\'u (Barcelona), Spain,
              and Institut d'Estudis Espacials de Catalunya,
              Ed. Nexus-201, C/ Gran Capit\`a 2-4, E-08034 Barcelona, Spain.
         \and
              Institut de Ci\`encies de l'Espai, CSIC, Barcelona, Spain,
              and Institut d'Estudis Espacials de Catalunya,
              Ed. Nexus-201, C/ Gran Capit\`a 2-4, E-08034 Barcelona, Spain.
             }

   \date{Received November 15, 2001; accepted --}

   \authorrunning{S. Goriely et al.}
   \titlerunning{He-detonation in CO white dwarfs}

\abstract{ He-accreting white dwarfs with sub-Chandrasekhar mass
 are revisited. The impact of the use of an extended reaction network on the
predicted energy production and characteristics of the detonating layers is studied.  
It is shown that the
considered scenario can be the site of an $\alpha$p-process combined with a
p-process and with a variant of the rp-process we refer to as the pn-process. We define
the conditions under which the derived distribution of the abundances of the 
p-nuclides in the ejecta, including the puzzling light Mo and Ru isotopes, mimics the 
solar-system one. 
}
  
 \maketitle

%

\section{Introduction}

The outcome of He-accreting sub-Chandrasekhar white dwarfs (WD) has deserved a special 
attention since the early 80's (e.g. Nomoto 1982, Woosley et al. 1986). 
Iben \& Tutukov (1991) have investigated the evolution of a close binary system
leading to the formation of a compact CO WD accreting He from a nondegenerate
low-mass companion. Limongi \& Tornambe (1991) concluded that such
systems could in some conditions lead to explosive
phenomena. Their relatively high estimated frequency, around 0.01 y$^{-1}$
(Iben \& Tutukov 1991), have drawn attention to their possible connection with
the progenitors of Type Ia supernovae.

He-accreting CO WDs are not viewed today as the most likely candidates for such
explosions (e.g. H\"oflich \& Khokhlov 1996, Hillebrandt \& Niemeyer 2000,
Branch 2001), but they might well be responsible for some 
special types of events. In fact, some one-dimensional calculations (e.g. 
Woosley \& Weaver 1994 and references therein) have concluded that 
He-detonations
on the considered WDs could well be identified as peculiar supernovae,
characterized by rapidly declining light curves with lower maximum luminosities
than those reached by C-deflagration Chandrasekhar-mass WDs.
These properties are reminiscent of subluminous supernovae like SN 1991bg.
Multidimensional simulations have confirmed the onset of the He-detonation, but 
have revealed significant differences in the central C-ignition which may be
triggered by the He-detonation (Livne \& Arnett 1995, Garcia-Senz et al. 1999).

This Letter limits its focus to some aspects of the
surface He detonation which have not deserved much attention up to now. More
precisely, we want to test the classical practice of calculating the energy production
and associated nucleosynthesis through a nuclear reaction network made of
$(\alpha,\gamma)$ captures. This approach is 
obviously unable to treat the production or captures of protons and neutrons 
in the detonating layers as well as their impact on the energetics and the
nucleosynthesis of the He detonation. Concomitantly, we present the first detailed
calculation of the synthesis of the nuclides heavier than the iron peak in the considered
He detonation. These problems are tackled in the framework of a 1-D model of 
the He detonation, the details of which are presented in
Sect.~2. Section 3 discusses the impact of the use of an extended reaction network on the
predicted energy production and characteristics of the detonating layers.  The composition
of the ejected material is analyzed in Sect.~4. We demonstrate that the considered scenario
can be the site of an $\alpha$p-process combined with a p-process and
with a variant of the rp-process we refer to as the pn-process. We define the
conditions under which the derived distribution of the abundances of the 
p-nuclides mimics the solar-system one. 
Conclusions are drawn in Sect.~5. 

\section{The He-detonation model}

We consider a non-rotating 0.8 $\msun$ CO-WD made of 30\% C and 70\% O by mass (Salaris et
al. 1997). It is stabilized and cooled to an initial luminosity $L_{wd} = 0.01 L_\odot$,
the central density and temperature being $\rho_c = 1.13\times10^7$ g~cm$^{-3}$, and 
$T_c = 1.14\times 10^7$ K. The accreted matter is He-rich (X(${}^4$He)=0.98), and is
assumed to contain traces of other nuclides: X($^{12}$C)=$5\times 10^{-3}$,
X($^{16}$O)=$5\times 10^{-3}$ and X($^{14}$N)=$10^{-2}$ [$X(i)$ is the mass fraction of
nuclide $i$].  As pointed out by
 Woosley \& Weaver (1994), the consideration of an initial non-zero
${}^{14}$N amount is  critical.
The \chem{14}{N} mass fraction adopted here
results from the burning of H in the companion assumed to be of typical Pop I
composition, but its precise value is considered by Woosley and Weaver
(1994) not to be critical for the He detonation outcome.  The accretion rate on the
CO-WD is adopted equal to $3.5\times 10^{-8} \msuny$, in agreement with the values
reported by Limongi \& Tornamb\`e (1991). The classical assumption
is also made that the accreted material has the same specific entropy as the outermost WD
shells. In these conditions, a thick He-rich envelope forms on the WD surface.

The evolution of this envelope is followed with a modified version of the code SHIVA
described by Jos\'e \& Hernanz (1998). It is a spherically symmetric, implicit,
hydrodynamic code in Lagrangian formulation which has been used extensively for the
modeling of classical nova outbursts. For the simulation reported here, a fine Lagrangian
mass grid of 400 shells is adopted. Shell masses range from $10^{-7} \msun$
for the innermost shells to $10^{-4} \msun$ for the outer layers.
The adopted nuclear reaction network is discussed in Sect.~3.

A thermonuclear runaway develops near the base of the He envelope when about
0.18~$\msun$ has been accreted. At this point, the density reaches a critical value
of about $10^6 \g~\cm^{-3}$ allowing the transformation  
\chem{14}{N}$(e^-,\nu)$\chem{14}{C}. When the temperature gets high enough, the resulting
\chem{14}{C} transforms into \chem{18}{O} through \reac{14}{C}{\alpha}{\gamma}{18}{O},
this $\alpha$-particle consumption channel competing successfully with the
$3\alpha$-reaction in the relevant temperature and density regimes (Hashimoto et al.
1986; also Piersanti et al. 2001). The associated energy release triggers the
detonation of He. More precisely, two shock waves propagate inward and outward from
the He-ignition shell. The outward-moving He-detonation wave heats the matter to
temperatures around $3\times 10^9 \K$. The expansion velocities  range from
$1000~\km~\s^{-1}$ in the vicinity of the He-ignition shell to more than
$20000~\km~\s^{-1}$ when the front reaches the WD surface, which is achieved in only
$0.176~\s$.  The whole envelope is ejected into the interstellar medium.
                        
The ingoing compressional wave pushes the inner WD material to velocities of nearly 
$\sim 3000~\km~\s^{-1}$. Its temperature, however does not exceed
$5\times 10^8 \K$, which does not allow to trigger the burning of carbon. The WD centre
is reached by the compressional wave after 0.7 s. As a result, carbon ignites, and
a second detonation develops near the centre. 
The high temperatures encountered during the C-deflagration ($T_9 \ga 4$), as well as the
initial composition of the CO WD lead mainly to the production of iron-peak nuclei
and are not expected to affect the nucleosynthesis of the p-nuclei.  For this reason, the
calculations do not follow the evolution of the inner part after C-ignition, so that only
the fate of the He-detonating envelope is modeled here. 

\section{The energetics of the He detonation}

\begin{figure}
\centerline{\epsfig{figure=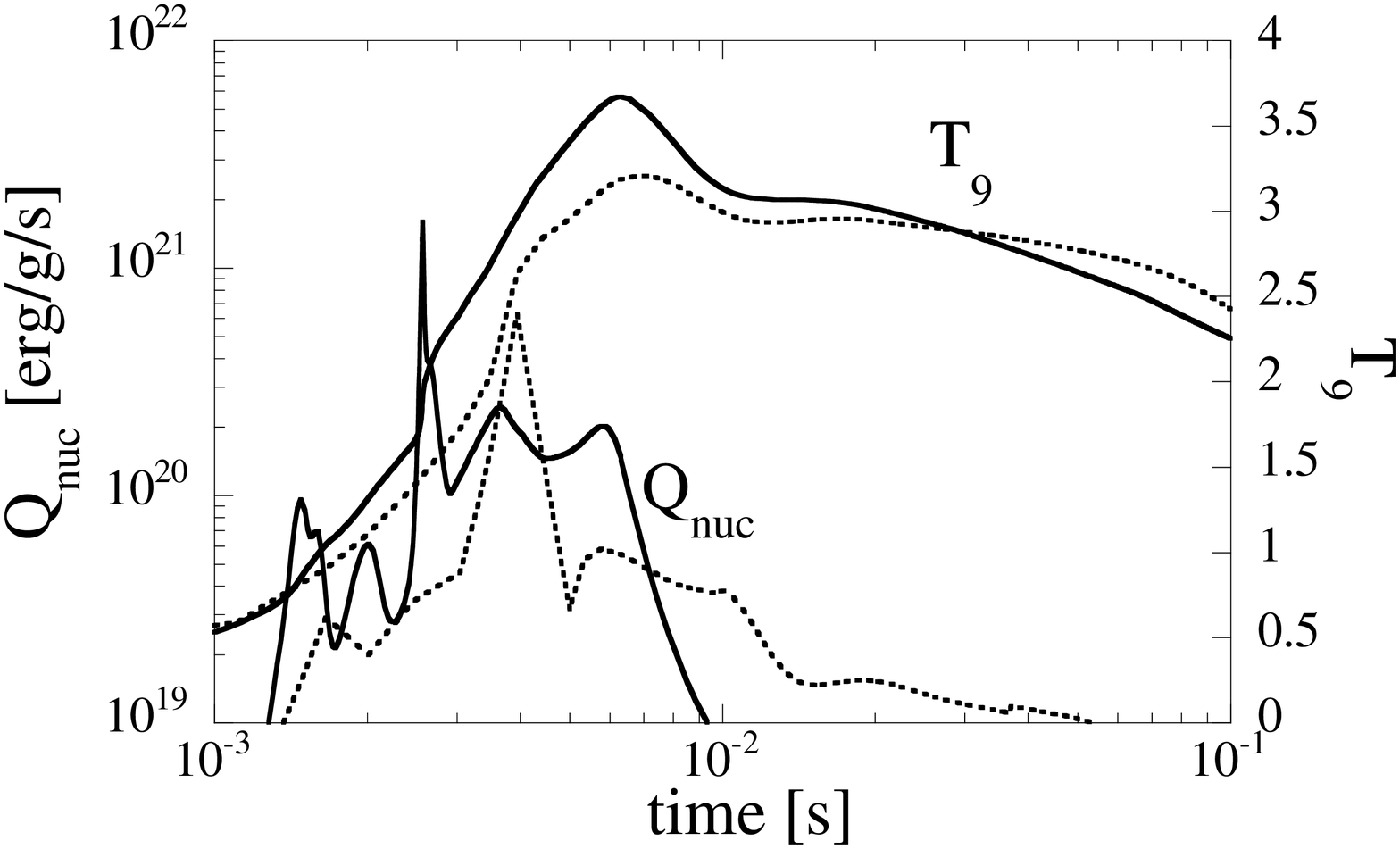,height=5.5cm,width=8cm}}
\caption{Comparison of the evolution of the rate of nuclear energy production and of
temperature predicted with the reduced (dashed lines) and with the extended network
(solid lines) in a layer located about $5\times 10^{-3}~\msun$ above the base of the
burning shell. }
 \label{F1}
\end{figure}

The estimate of the energy released by the He detonation is generally derived from the use
of a limited network of some 50 nuclear reactions and 26 nuclides. In addition to the
$(\alpha,\gamma)$-chains from He to Ni, it is made of the
${}^{14}$N($e^-,\nu){}^{14}$C($\alpha,\gamma){}^{18}$O and the C+C, C+O and O+O
reactions.  Reverse photodisintegrations are also included. 
Post-processing calculations based on a full network to be described
in Sect.~4 demonstrate that the detonation produces substantial amounts of neutrons and
protons through $(\alpha,n)$ and $(\alpha,p)$ reactions. They induce
nucleon capture reactions that modify substantially the nuclear flow predicted by the
limited network, and concomitantly the energetics of the detonation and the associated
nucleosynthesis.  For these reasons, an extended nuclear reaction network inspired by the
post-processing calculations has been implemented in the hydrodynamic simulations. It
includes 188 nuclides up to $^{68}$Zn linked through a net of 571 nuclear reactions. These
reactions are selected on grounds of the fact that they contribute to more than one
per mil of the total energy generated at any timestep. They comprise  neutron, proton and
$\alpha$-captures, as well as photodisintegrations and $\beta$-decays. Although the new
calculations are extremely time consuming, they are considered to be essential for a
reliable determination of the energetics of the He detonation and of the thermodynamics of
the He shells. In fact, the pre-explosion evolution at $T<10^8$ K is not affected by the
network extension. Differences are found at temperatures in excess of $10^9$ K, where 
a large number of reactions neglected in the reduced networks are responsible for an
increase of the energy deposited in the envelope. In our simulations, the reduced
network leads to a peak temperature of the He-burning shell of $3.38\times 10^9$ K,
with a maximum rate of energy production of $Q_{\rm nuc} = 2.2\times 10^{21}$ erg/g/s,
while a peak temperature of $3.83\times 10^9$ K  and $Q_{\rm nuc} = 4.5\times 10^{22}$
erg/g/s are obtained with the extended network. The He-detonation wave hits the WD
surface a bit earlier (0.166 s) than with the reduced network (0.176 s).  As an example,
we display in Fig.~\ref{F1} the quite significant differences in the evolution of $T$
and $Q_{\rm nuc}$ predicted with the two networks for a layer located about
$5\times 10^{-3}~\msun$ above the base of the He-burning shell. In fact,  the considered
layer is seen to  experience a unique energy burst due to the $\alpha$-chain of the
reduced network (involving solely
$N=Z$ nuclei), and in particular to
 $^{20}$Ne$(\alpha,\gamma)^{24}$Mg$(\alpha,\gamma)^{28}$Si. The extended network leads
to an initial double energy burst, the first one being due to
$^{14}$C$(\alpha,\gamma)^{18}$O, the second one resulting from radiative neutron
captures on the most abundant species. The reactions $^{18}$O$(\alpha,n)^{21}$Ne,
followed by  
$^{22}$Ne$(\alpha,n)^{25}$Mg and $^{26}$Mg$(\alpha,n)^{29}$Si are indeed responsible
for a  neutron density as high as $N_n \simeq 10^{22}~{\rm cm}^{-3}$. At later times,
the major energy burst results from the main $\alpha$-chain burning. It develops
earlier than the peak obtained with the reduced network, this time shift being due to the
faster temperature increase predicted with the extended network. It is followed by a
secondary $Q_{\rm nuc}$ peak resulting from the capture of protons produced by $(\alpha,p)$
reactions on $Z\simeq N$ nuclei, and in particular on $^{44}$Ti, the mass fraction of which
reaches about $5\times 10^{-3}$. At the peak temperature of $T_9\simeq 3.6$ reached in
this
 layer, the proton mass fraction amounts to $X_p\sim 6\times 10^{-3}$. 

\section{The nucleosynthesis in the He detonation}

The composition of the $0.18 \msun$ of ejected envelope is evaluated by a post-processing
nucleosynthesis calculation based on the temperature and density profiles derived from the
extended-network simulation described in Sect.~3.  A full network including
some 50000 reactions on about 4000 nuclides up to Po and lying between the proton and
neutron drip lines is solved for each of the 100 envelope layers. All experimental and
theoretical reaction and $\beta$-decay rates are taken from the Nuclear Network Generator
of the Brussels Library (Jorissen \& Goriely, 2001). This network is also the one used
to define the minimum network that had to be implemented in the hydrodynamic simulations to
calculate the nuclear energy production in each layer of the model (Sect.~3). The
initial envelope composition is described in Sect.~2 for the light nuclei and assumed to
be solar above Ne.

As already explained in Sect.~3, the pattern of nuclear reactions developing
during the explosion is rather complex. Initially, a high neutron irradiation 
(neutron densities $N_n \simeq 10^{22}-10^{23}~{\rm cm}^{-3}$) originating from
$^{18}$O$(\alpha,n)^{21}$Ne, and later from $^{22}$Ne$(\alpha,n)^{25}$Mg and
$^{26}$Mg$(\alpha,n)^{29}$Si, drives most of the flow to the neutron-rich side of the
nuclear chart. A weak r-process ensues. However, the increase of temperature above $T_9
\ga 1.5$  induces fast photodisintegrations driving the
matter back to the valley of $\beta$-stability, and even to its neutron-deficient side.
From this point on, two major nucleosynthesis processes take place.

In the layers with peak
temperatures $2 \la T_9 \la 3$, a typical p-process is found reponsible for the production
of the stable p-nuclides (Rayet et al. 1995). In these conditions, the nuclear flow is
dominated by ($\gamma$,n), ($\gamma$,p) or ($\gamma$,$\alpha$) photodisintegrations,
complemented with mainly some neutron captures. For layers with peak temperatures $T_9 \ga
3$, large amounts of $^{40}$Ca and $^{44}$Ti are produced by radiative $\alpha$-captures.
Further
$\alpha$-captures proceed through ($\alpha$,p) reactions, so that a so-called
$\alpha$p-process develops, the resulting proton mass fraction reaches about $X_p =
6\times 10^{-3}$. In the considered hot environment, these protons are rapidly captured
to produce heavier and heavier neutron-deficient species, making up a kind of
`proton-poor rp-process', in view of the much lower proton concentrations than in the
`classical' rp-process. In this process, some nuclei are produced with proton separation
energies that are small enough to experience $(\gamma,p)$ photodisintegrations which
slow down the nuclear flow. However,
$(n,p)$ reactions made possible by the high neutron density (at this stage, $N_n\simeq
10^{19}~{\rm cm}^{-3}$) revive the flow towards higher-mass nuclei by new p-captures. One
might thus talk about a `proton-poor neutron-boosted rp-process', which we coin the
pn-process. The nuclear flow associated with this variant of the rp-process lies much
further away from the proton-drip line than in the classical rp-process. This results from
the lower proton and non-zero neutron concentrations encountered in the He detonation. A
detailed discussion of this pn-process nuclear flow and of the associated nuclear physics
uncertainties will be presented elsewhere.

\begin{figure}
\centerline{\epsfig{figure=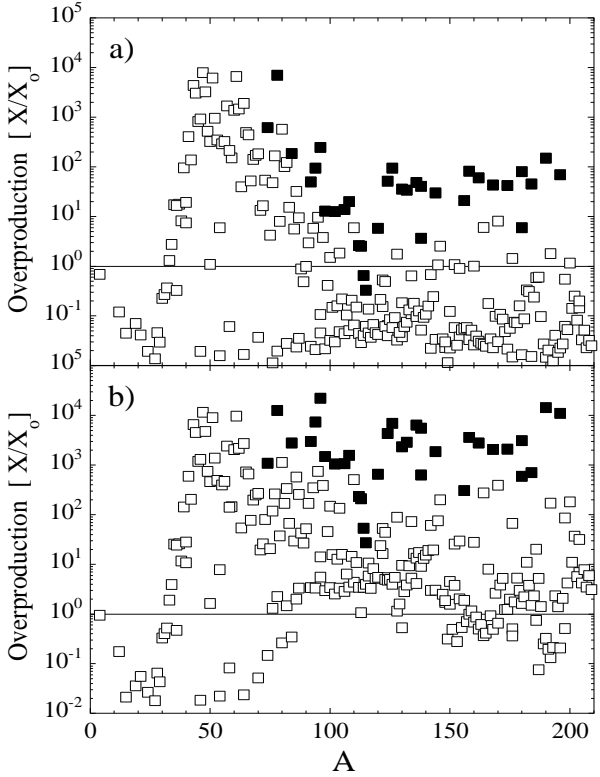,height=10.4cm,width=8cm}}
\caption{(a) Final composition of the ejected envelope as a function of the mass number
$A$. Full symbols denote the p-nuclides; (b) Same as (a), but the initial abundances of
the s-nuclides is assumed to be 100 times solar.}
 \label{F2}
\end{figure}

The final envelope composition is displayed in Fig.~\ref{F2}a. As a new nucleosynthesis
prediction associated with the He detonation, we note that almost all the p-nuclei are
overproduced in solar proportions within a factor of 3 as a combined result of the
p- and pn-processes. 
This includes the puzzling Mo and Ru p-isotopes (Rayet
et al. 1995, Costa et al. 2000) which are efficiently produced at peak temperatures $T_9
\ga 3.5$. The lighter Se, Kr and Sr p-isotopes are synthesized in layers heated to $3\la
T_9 \la 3.5$, \chem{78}{Kr} being the most abundantly produced in these conditions.
The high sensitivity to temperature of the production of the $A < 100$ p-nuclei makes the
correct description of the corresponding layers (and thus the use of a suitably extended
reaction network) mandatory.  

Fig.~\ref{F2}(a) also makes evident that the Ca-to-Fe nuclei are overabundant
with respect to the p-nuclei but \chem{78}{Kr} by a factor of about 100, which implies that
the considered He detonation is not an efficient scenario for the production of the bulk
solar-system p-nuclides. In order to cure this problem, one may envision enhancing the
initial abundance of the s-nuclides, which are the seeds for the p-process.
Fig.~\ref{F2}(b) shows that an increase by a factor 100 of the s-nuclide abundances
over their solar values makes the overproduction of a substantial variety of p-nuclides
comparable to the one of \chem{78}{Kr} and of the Ca-to-Fe nuclei. The factor 100
enhancement would have to be increased somewhat if the material processed in the core of
the CO-WD by C-detonation were ejected along with the envelope. At this point, one
essential question concerns the plausibility of the required s-nuclide enhancement. We do
not have any definite answer to this key question.  The required s-process enrichment of
the accreting WD might result from its past AGB history if indeed some of its outer
s-process enriched layers could be mixed convectively (or due to rotationnal effects) 
with part at least of the accreted He-rich layers before the detonation. Alternatively, the
He-rich matter accreted by the CO-WD could be (or become) enriched in s-process elements.
Such speculations (e.g Iben \& Tutukov 1991) need to be confirmed by detailed simulations.

\section{Conclusion}
 
This Letter presents the first instance of a clear possibility for $\alpha$p processed
material to be ejected into the interstellar medium. In previously proposed sites, like
accreting neutron stars associated to  X-ray bursts or accretion disks around
black holes (e.g. Schatz et al. 1998), such an ejection is indeed far from being
demonstrated. Of course, the global contribution of He-detonating CO-WD to the galactic
nuclidic content remains uncertain, but there is reasonable hope for it not to be
negligible in view of the predicted frequency of about 0.01 per year for these events. We
also find that this galactically `fertile' $\alpha$p process is accompanied with an
efficient p-process and triggers a variant of the rp-process, the pn-process, which develops
in the presence of neutrons and with less protons than the classical rp-process. Most of
the p-nuclides, including the puzzling light Mo and Ru isotopes, are found to be
co-produced in these conditions in relative quantities close to solar. Unfortunately, they
are underproduced (except \chem{78}{Kr}) with respect to the Ca-to-Fe species. The price
to pay in order to avoid this difficulty is an increase of the abundances of the seed
s-nuclei by a factor of about 100 over their solar values. The astrophysical plausibility  
of this enhancement remains to be scrutinized in detail, in particular by
studying the impact of rotationally induced mixing.

In spite of this problem, we consider that the results presented here are encouraging
enough for justifying an extension of our calculations to other situations involving
CO-WD of different masses and accretion rates. These additional simulations will be
presented elsewhere, along with a detailed discussion of the characteristics and nuclear
physics uncertainties of the associated $\alpha$p, pn- and p-process flows.

\acknowledgements{S.G. and M.R. are FNRS research associates. Work partially
 supported by the Spanish MCYT (J.J. and M.H.).}

\end{document}